\documentclass[twocolumn,showpacs,amssymb,amsmath,nobibnotes,prl]{revtex4}

\usepackage{graphicx}
\usepackage{xcolor}
\usepackage{bm}%

\begin{document}

\title{{Explosive  Magnetorotational  Instability in Keplerian Disks}}

\author{Yu. Shtemler}%
  \email{shtemler@bgu.ac.il}
\author{E. Liverts}%
 \email{eliverts@bgu.ac.il}

\author{M. Mond}
 \email{mond@bgu.ac.il}
\affiliation{
Department of Mechanical
Engineering,\\ Ben-Gurion University of the Negev\\ Beer-Sheva 84105,
Israel
}

\date{\today}

\begin{abstract}
Differentially rotating disks under the effect of
axial
  magnetic field are prone to
a  nonlinear explosive magnetorotational instability (EMRI).
  The dynamic  equations that govern the temporal evolution of the amplitudes of   three
  weakly-detuned resonantly
  interacting modes are derived. As distinct from exponential growth in the strict resonance triads
 EMRI
  occurs due to the resonant  interactions of a
 MRI mode
 with stable Alfv\'{e}n-Coriolis and magnetosonic modes. Numerical solutions of the dynamic  equations
   for  amplitudes of a triad
   indicate that
   two types of perturbations behavior can be excited for resonance conditions: (i)
  EMRI
    which leads to infinite values of the three amplitudes within a finite time, and (ii) bounded irregular oscillations of all three amplitudes.
 Asymptotic explicit
 solutions of the dynamic  equations are obtained
for
EMRI regimes
   and are shown to match the numerical solutions near the explosion time.
\end{abstract}
\pacs{PACS 98.62.Mw }

\maketitle

\section{\label{sec:level1}I. Introduction}
The magnetorotational instability (MRI, \cite{Velikho1959} \nocite{Chandra1959}  -- {\cite{BALBUS:1991sp}) is believed to play a key role in the angular momentum transfer in accretion disks, and has been thoroughly investigated through linear analysis as well as nonlinear magnetohydrodynamic (MHD) simulations under a wide range of conditions and applications. First attempts to achieve analytical insight into the nonlinear evolution of the MRI focused on the dissipative saturation of the instability (\cite{Knobloch:2005la}-- \cite{Umurhan:2007eu}) in environments that are characteristic of laboratory experiments.

Recently it has been suggested that the mechanism of dissipationless wave interaction plays an important role in the nonlinear development of the MRI in astrophysical disks (\cite{Liverts:2012fr} \nocite{Shtemler:2013qv} \nocite{Shtemler:2014uq}--\cite{Shtemler:2011bh}).
 According to one such scenario the MRI forms a triad of interacting modes with a stable slow or fast Alfv{\'e}n-Coriolis (AC) mode  and a stable magnetosonic (MS) mode. This is a generalisation and adaptation to the thin, rotating, axially stratified disks of the well known three-wave interaction in static, homogeneous and infinite plasmas and fluids (\cite{Galeev:1961}-- \cite{SagdeevandGaleev}).
  The saturation of the MRI by non-resonant excitation of a MS wave, as well as by resonantly exciting two small-amplitude linearly stable modes have been investigated
   for astrophysical disks
     (\cite{Liverts:2012fr} \nocite{Shtemler:2013qv} \nocite{Shtemler:2014uq} -- \cite{Shtemler:2011bh}).
It has been shown that in the  resonance case
 with zero frequency  mismatch
 the two linearly stable modes
are either bounded or
grow exponentially 
due to the nonlinear coupling to the saturated MRI.
 In the case of zero frequency mismatch is named hereafter
strict resonance in order to distinct it on the general resonance case of small nonzero frequency mismatch.
 As is also well known  for
 different dispersive plasma media,
  nonlinear interactions of  three resonant modes
 may give rise to explosive instability (EI),  under which the amplitudes of the triad modes
grow to infinity in finite time
(\cite{Weiland} \nocite{Craik} \nocite{rosenbluthetal}
  \nocite{Oraevski:1973}  \nocite{Fukai:1969} \nocite{Dum and Ott:1971} \nocite{Davydova et al:1978}--\cite{Pavlenko:1978}).

The  current work is focussed  on the evolution of resonant triads  in thin Keplerian disks.
 In particular, it is shown that the exponential nonlinear instability of the strict resonance triads is replaced by  explosive MRI
 (EMRI)
 in resonance conditions with small non zero frequency mismatch.

\section{\label{sec:level2}II. The Physical Model}
Nonlinear  evolution of the  axisymmetric MRI in spatially stratified rotating plasma  is studied for  Keplerian disks threaded by a poloidal magnetic field.
 The thin disk approximation
 (\cite{REGEV:1983pb}--
 \cite{Shtemler:2007ly}) is applied to the MHD equations by ignoring viscosity, electrical resistivity, and radiation effects.
    This  approximation  developed    in \cite{Shtemler:2013qv}--  \cite{Shtemler:2014uq} for strict resonance interactions in thin Keplerian disks  is  applied in the present study to detuned resonance triads.  Hereby the main results of these studies are summarized in that section:

\subsection{\label{sec:level2A}A. Governing relations}
Assuming the disk to be vertically isothermal yields the
  following dimensionless dynamic  system:
 \begin{equation}
\frac{ D \bf {V} }{Dt} = -\bar{C}^2_S\frac{\nabla n}{n}+\frac{1}{\beta_* } \frac{\bf{j}
\times {\bf{B}}}{n}- \epsilon^{-3} \nabla \Phi,\,\,\,\,
\label{Mom}
\end{equation}
\begin{equation}
\frac{\partial n} {\partial t}
               + \nabla \cdot(n {\bf{V}} ) =0,\\
\label{Mas}
\end{equation}
\begin{equation}
\frac{\partial {\bf{B} }} {\partial t}+
 \nabla \times
 {\bf{E}}=0,\,\,\, {\bf {E} }=  -  {\bf{V}} \times {\bf{B}},
\label{Magn}
\end{equation}
\begin{equation}
 {\bf {j} }= \nabla   \times {\bf{B}},\,\,\,\,\,\,\,\,\,\nabla \cdot {\bf{B}}=0,
\label{Cur}
\end{equation}
\begin{equation}
B_r=B_\theta=0,\,\,\,n=0\,\,\,\mbox{for} \,\,\, z=\pm\infty.
\label{BC}
\end{equation}
The assumption of
vertically isothermal disk
 implies that  $\nabla P=\bar{C}^2_S\nabla n$,
where $\bar{C}^2_S =
 \bar{T}(r)$ is  the dimensionless sound speed;  $T=\bar{T}(r)$  is the steady-state background temperature; $P(r,z,t)$, $n(r,z,t)$,  $\bf{V}$   and $\Phi(r,z)$ are the total plasma pressure, number density, plasma velocity
  and
 gravitational potential due to the central body;
  $\{r,\theta,z\}$   are the  cylindrical coordinates;
$t$ is time; $D/Dt=\partial/\partial
t+(\bf{V}\cdot\nabla)$  is the material derivative;  $\bf{B}$, $\bf{j}$ and $\bf{E}$ are the magnetic field,    current density and
  electric field, respectively.

 A MHD  model of dissipationless quasi-neutral plasmas is characterized  by three dimensional velocities calculated at the characteristic  radius, $r_*$, namely, the Keplerian rotation,   $V_{K*}=r_*\Omega_*\equiv\sqrt{G\, M_c/r_*}$, the sound velocity,  $C_{S*}=\sqrt{T_*/m_i}$, and the Alfv\'en velocity,  $V_{A*}=B_*/\sqrt{4\pi m_i n_i}$, which  produce two dimensionless parameters: the plasma beta $\beta_*$ and the  Mach number $M_*$ that is equals to the inverse small disk aspect ratio $\epsilon=H_*/r_*$
  \begin{equation}
\beta_*=4\pi\frac{P_*}{B_{*}^2}\equiv \frac{C_{S*}^2}{V_{A*}^2},\,\,\, M_*=\frac{V_{K*}} {C_{S*}}\equiv \epsilon^{-1}
.\,\,\,
\label{BetM}
\end{equation}
 The  dimensional values $M_c$, $G$, $m_i$, $n_i$ and $T_*$, $B_*$, $H_*=C_{S*}/\Omega_*$ are as follows: the mass of the central body, the gravitational constant, the ion mass and number density, and the temperature, poloidal magnetic field  and the disk height calculated at $r=r_*$.

\subsection{\label{sec:level2B}B. Steady-state}
The smallness of   $\epsilon$   makes it useful to define a slow radial coordinate, and
 in addition
  to leading order in   $\epsilon$, it is convenient to replace the  independent variables by the following
self-similar
 quantities:
\begin{equation}
\xi=\epsilon r\sim
\epsilon^0
, \,\,
\tau= \bar{\Omega}(\epsilon r)t\sim \epsilon^0,\,\,
\eta=z/\bar{H}(\epsilon r)\sim\epsilon^0.
\label{NewV}
\end{equation}
The radial and axial coordinates  $\xi$  and  $\eta$ are  stretched in compliance with assumption of small disk aspect ratio;
$\bar{H}(\xi)=\bar{C}_s(\xi)/\bar{\Omega}(\xi)$ is the semi thickness of the disk.
 The  Keplerian steady-state configuration is described to leading order in $\epsilon$  by the equilibrium conditions
in the radial and axial directions as follows:
$$
\Phi(r,z)=-\frac{1}{\sqrt{r^2+z^2}}\equiv
-\epsilon\bar{\Phi}(\xi)+\frac{\epsilon^3}{2}\bar{C}_S^2(\xi)\bar{\psi}(\eta)
+O(\epsilon^5),
\,\,
$$
$$
\,\,\,\,\,\,\,\,\,\,\,
n(r,z)=\epsilon^0\bar{N}(\xi)\bar{n}(\eta)+O(\epsilon^2),
\,\,\,\,\,\,\,
\,\,\,\,\,\,\,\,\,\,\,\,\,\,\,\,\,\,\,\,\,\,\,\,\,\,\,\,
$$
$$
\,\,\,\,\,\,\,\,\,\,
V_r=V_z= 0,\,\,\,\,\,\,
V_\theta =\epsilon^{-1}\bar{V}(\xi)+O(\epsilon),
\,\,\,\,\,\,\,
\,\,\,\,\,\,\,\,\,\,\,\,\,\,\,\,\,\,\,\,\,\,\,\,\,\,\,\,
$$
\begin{equation}
\,\,\,\,\,\,\,\,\,\,
B_r=B_\theta=0,
\,\,\,B_z=\epsilon^0\bar{B}(\xi),\,\,\,
\,\,\,\,\,\,\,\,\,\,\,\,\,\,\,\,\,\,\,\,\,\,\,\,\,\,\,\,\,\,\,\,\,\,\,
\label{StGauge}
\end{equation}
where  $\bar{C}_S(\xi)$,  $\bar{B}(\xi)$, $\bar{N}(\xi)$  are arbitrary functions;\\
$
\bar{V}(\xi)\equiv \xi\bar{\Omega}(\xi)= \frac{1}{\sqrt{\xi}},
\bar{\Phi}(\xi)=\frac{1}{\xi},
\bar{\psi}(\eta)=\eta^2,\,
\bar{n}(\eta)=e^{-\eta ^2/2}.
$

\subsection{\label{sec:level2C}C. Nonlinear  perturbations}
Both steady-state equilibrium as well as the perturbed variables are scaled with well-defined powers of the small parameter  $\epsilon$:
\begin{equation}
F(r,z,t)=\epsilon^{\bar{S}}\bar{F}(\xi,\eta;\epsilon)+\epsilon^{S'} F'(\xi,\eta,\tau;\epsilon).
\label{Gauge}
\end{equation}
Here  $F$ stands for any dependent variable, the bar and the prime denote equilibrium and perturbed variables, respectively, which are characterized by gauge functions $\epsilon^{\bar{S}}$  and  $\epsilon^{S'}$. The various values of $\bar{S}$  are determined above in eqs. (\ref{StGauge}), while the values of $S'$ are
obtained
 to be zero for all the dependent variables except for  the perturbed axial magnetic field for which  $S'=1$.

The perturbed dependent variables can now be
  scaled with the following slow-radius varying functions:
\begin{equation}
{\bf{v}}=\bf{V}'/\bar{C}_S(\xi),\,\,\,\
\nu=n'/\bar{N}(\xi),\,\,\,\,
{\bf{b}}=\bf{B}'/\bar{B}(\xi).
\label{Scale}
\end{equation}
Then inserting Eqs. (\ref{NewV}) - (\ref{Scale}) in Eqs. (\ref{Mom})-(\ref{BC}), keeping the leading order terms in $\epsilon$ and eliminating $v_r$ and $v_\theta$ yield the following  dimensionless dynamic  equations:
\begin{equation}
\frac{\partial^2 b_r}{\partial \tau^2}-2\frac{\partial b_\theta}{\partial \tau}
-\frac{1}{\beta(\xi)}\frac{\partial }{\partial \eta}
\big{[}\frac{1}{\bar{n}(\eta)}\frac{\partial b_r}{\partial \eta}\big{]}-3 b_r=
N_r,
\label{br}
\end{equation}
\begin{equation}
\frac{\partial^2 b_\theta}{\partial \tau^2}+2\frac{\partial b_r}{\partial \tau}
-\frac{1}{\beta(\xi)}\frac{\partial }{\partial \eta}
\big{[}\frac{1}{\bar{n}(\eta)}\frac{\partial b_\theta}{\partial \eta}\big{]}=
N_\theta,
\label{bThet}
\end{equation}
\begin{equation}
\frac{\partial^2 \nu}{\partial \tau^2}
-\frac{\partial }{\partial \eta}
\big{[}\bar{n}(\eta)\frac{\partial }{\partial \eta}
\big{(}\frac{\nu}{\bar{n}(\eta)}\big{)}
\big{]}=
N,
\label{nu}
\end{equation}
\begin{equation}
b_r=b_\theta=0,\,\,\,\,\nu=0\,\,\,\,\,\mbox{for}\,\,\,\,\,\eta=\pm\infty.
\label{BCsc}
\end{equation}
Explicit expressions for the
 nonlinear righthand sides $N_r$, $N_\theta$ and $N$ in eqs. (\ref{br}) -(\ref{nu}) as well as the equations for $v_r$, $v_\theta$  and $v_z$ that relate them to $b_r$, $b_\theta$ and $\nu$ are presented in \cite{Shtemler:2014uq}. The equation for the perturbed axial magnetic field decouples from the rest of the equations and drops out from the governing system. To leading order in $\epsilon$ the radial derivatives also drop from the resulting system  without approximation of frozen radial variable, and the radial dependence enters only through  the local plasma beta:
 $\beta(\xi)=\beta_* \bar{N}(\xi)\bar{C}_s^2(\xi)/\bar{B}^2(\xi)$.

\subsection{\label{sec:level2D}D. Linear  perturbations}
 Modifying the  mass density profile to $\bar{n}(\eta) =\mathrm{sech} ^2\eta$ enables the analytical solution of the linearized set of equations (\ref{br}) -(\ref{BCsc}) for small perturbations.  The resulting linear eigenmodes are
 decoupled into two families of the AC and MS modes
(\cite{Shtemler:2014uq} and references therein).
 The eigenfunctions of  AC modes
 for the in-plane perturbed velocities
 are expressed in terms of the Legendre polynomials
    $P_k(\zeta)$ ($k=1,2,\dots$) of  $\zeta=\tanh(\eta)$. The  AC family
    represents in-plane perturbations and forms a discrete spectrum whose eigenvalues are:
\begin{equation}
\omega_{k,l}=\pm
\sqrt{
\frac{\beta+6\beta_k+l\sqrt{(\beta+6\beta_k)^2-36\beta_k(\beta_k-\beta)}}{2\beta}.
}
\label{EigenFreq}
\end{equation}
  Here $\omega_{k,l}$ are labeled by two integers: $-\infty <k<\infty$
     that  plays the role of the axial wave number, and
   $l=-1(+1)$ that represents the slow (fast) AC modes.
 The fast AC modes are stable while the slow AC modes may become unstable. The number of unstable slow AC modes is determined by the local plasma beta $\beta(\xi)$.
  Thus, the threshold for exciting $k$ unstable modes is given by $\beta_k=k(k+1)/3, k=1,2,\ldots$. It is those unstable slow AC modes that constitute the MRI whose eigenvalues are given by $\omega _{m,-1}=i\gamma, \;m=1, \dots,k$. Of particular importance
  is the fact that for $\beta=\beta _k$, $\gamma _k=0$ is a double root of the dispersion equation.
  The second family of eigen-oscillations in thin Keplerian disks is the vertical MS modes. The latter are stable, possess a continuous spectrum, and their eigenfunctions
    are expressed in terms of  special functions \cite{Liverts:2012fr}.
  The  AC and  MS  families of the linear eigenmodes are the building blocks of the present nonlinear analysis be unfolded in the next sections.

\section{\label{sec:level3}III. Nonlinear dynamics of detuned triads}
\subsection{\label{sec:level2A}A. Scenario of three-mode nonlinear interaction}
The scenario that is introduced in the current work is a generalization of the mechanism described in \cite{Shtemler:2013qv}--  \cite{Shtemler:2014uq}   for the strict  resonance triads: A $\beta$ value {\it{slightly}} above the first threshold
 $\beta_1=2/3$
for MRI is considered. As a results there is only one unstable MRI mode, that is characterized by axial wave number $k=1$.
 A MRI eigenmode that is characterized by complex frequency
  $0+i\gamma $  forms a triad of interacting modes with a stable fast or slow AC mode  and a stable MS wave that are characterized by a real frequency $\omega _a\equiv\omega_{k,l}$ and $\omega _s$, respectively. The condition for the occurrence of  resonant interaction between the three eigenmodes is $\omega_s=\omega _a +\Delta \omega$, where the frequency mismatch $\Delta \omega$ is much smaller than each of the eigenfrequencies $\omega _a$ and $\omega _s$.
The resonance condition is easily satisfied due to the continuous nature of the MS spectrum. This  condition can be satisfied also if the Gaussian  mass distribution is considered. In that case the MS spectrum is discrete, however, as was pointed out in \cite{Shtemler:2014uq}, for any frequency $\omega _{k,l}$ of a stable AC mode, a corresponding eigenfrequency of the Gaussian MS spectrum may be found such that the resonance condition is satisfied.
The customary resonance condition on the axial  wave number is not needed here due to the axial stratification of the mass density; it is replaced by the solvability conditions of the higher orders boundary value problem in the axial coordinate.
 The resonance condition on the radial wave number drops out
in the  thin disk approximation according to which the radial derivatives are negligible.
   The three-wave interaction is a direct result of the influence of the perturbed in-plane magnetic pressure gradients on the acoustic modes, and the simultaneous axial convection of the AC modes by the acoustic perturbations.

The main goal of the current section is to derive a set of coupled ordinary differential equations that govern the evolution of the amplitudes of the modes that take part in the resonant interaction. During the linear stage those amplitudes are constants that are determined by the initial conditions. However as the resonant interaction gains in importance, the mutual interaction changes dramatically their time evolution. Thus, while the equations that govern the amplitudes of the stable AC and MS modes are similar to their first order classical counterparts \cite{SagdeevandGaleev},  the equation for the amplitude of the MRI is of second order, reflecting the multiplicity two of the eigenvalue at the threshold beta \cite{Shtemler:2014uq}.

Deriving the amplitude equations starts with observing that for small values of the growth rate $\gamma$
and frequency mismatch $\Delta \omega\sim \gamma$,
   the  resonant interactions are described by two distinct time scales:  a fast time ${\bar \tau}=t$, and a slow time ${\tilde \tau}=\gamma t$. Assuming further that the frequency mismatch $\Delta \omega$ is of order $\gamma$, each of the perturbations due to the AC modes ($f(\eta, t)$) as well as those due to the MS waves ($g(\eta, t)$) may be represented as a sum of zeroth and first harmonic terms in the fast time  (higher harmonics are neglected), each with an  amplitude varying with the slow time:
\begin{equation}
f(\eta ,t)=f_0(\eta ,{\tilde \tau})+[f_1(\eta ,{\tilde \tau})e^{-i{\omega _a\bar \tau}}+c.c],
\label{AC}
\end{equation}
where the subscripts $0$ or $1$ denote the harmonic number. A similar expression is written also for $g(\eta, t)$. Contributions to  $f_0$, the zeroth harmonic AC perturbations, come from the MRI, a non-resonant excitation of a zeroth harmonic MS wave, and the interaction of the stable fast AC and the MS eigenmodes, all of which may be represented  as:
\begin{eqnarray}
\nonumber
f_0(\eta ,{\tilde \tau})&=&A_0({\tilde \tau}) \Psi_0(\eta)+A_0({\tilde \tau})H_0({\tilde \tau})\psi_{0,0}(\eta)\\
&+&[A_1^*({\tilde \tau})H_1({\tilde \tau})\psi_{-1,1}(\eta)+c.c],
\label{zeroharmonic}
\end{eqnarray}
where $A_0$ is the real-valued amplitude of the MRI ($e^{\pm \tilde \tau}$ during the linear stage), $\Psi_0$ its linear eigenfunction, $A_1, H_1$ are the amplitudes of the stable AC and MS modes, respectively (constants during the linear stage), $H_0$ is the amplitude of the non-resonantly excited MS wave, and $\psi_{-1,1}, \psi_{0,0}$ are yet to be determined coupling functions. In a similar manner the expressions for $f_1,g_1$are given by:
\begin{equation}
f_1(\eta,{\tilde\tau})=A_1({\tilde\tau})\Psi_1(\eta)
+[A_0({\tilde \tau})H_1({\tilde \tau})\psi_{0,1}(\eta)+c.c.],
\label{firstharmonica}
\end{equation}
\begin{equation}
g_1(\eta,{\tilde \tau})=H_1({\tilde \tau})\Phi_1(\eta)
+[A_0({\tilde \tau})A_1({\tilde \tau})\phi_{0,1}(\eta)+c.c.],
\label{firstharmonicm}
\end{equation}
where $\Psi_1, \Phi_1$ are the eigenfunctions of the AC and MS modes, respectively, $\phi _{0,1}(\psi _{0,1})$ is the coupling function between the MRI and the AC(MS) mode. The first terms on the right hand sides of eqs. (\ref{zeroharmonic})-(\ref{firstharmonicm}) describe the three linear modes that participate in the  resonant interaction. The amplitude of the non resonantly driven MS wave may be shown to be $H_0({\tilde \tau})=A_0^2({\tilde \tau})$ \cite{Shtemler:2014uq}.

\subsection{\label{sec:level2B}B. Nonlinear equations for triad amplitudes}
The amplitudes $A_0, A_1$, and $H_1$ are the main players in the current work and deriving the set of ordinary differential equations that govern their slow-time dynamic  evolution is the main concern of this section.
  Thus, recalling the second order degeneracy of the MRI
 eigenfrequency
    at the threshold  beta eq. (\ref{EigenFreq})),
    the dynamic  equation for $A_0$ is necessarily of second order, that in view of its resonance interaction with the two other modes is conjectured to be given by:
\begin{equation}
\gamma ^2\frac{d^2 A_0}{d{\tilde \tau}^2}=\gamma ^2 A_0+\Gamma _{0,0}A_0^3+[\Gamma _{-1,1}A_1^*H_1e^{-i{\delta\omega\,\tilde \tau} }+c.c].
\label{amplitude}
\end{equation}
The corresponding first order equations for $A_1$ and $H_1$  are identical with their classical counterparts \cite{SagdeevandGaleev}:
\begin{equation}
\gamma \frac{dA_1}{d{\tilde \tau}}=i\Gamma _{0,1}A_0H_1e^{-i\delta\omega\, {\tilde \tau}}, \; \gamma \frac{dH_1}{d{\tilde \tau}}=i\Gamma _{1,0} A_0A_1e^{i\delta\omega\, {\tilde \tau}}.
\label{firstorder}
\end{equation}
Here $ \delta\, \omega= \Delta \omega/\gamma\sim\gamma^0$, and $\Gamma _{i,j}\sim\gamma^0$ are nonlinear coupling coefficients  yet to be determined that measure the interaction between the various waves.
  A clear hierarchy emerges from
  the principle of   least degeneracy of
eqs. (\ref{amplitude})-(\ref{firstorder}) according to  which
 the amplitude of the  MRI mode in thin Keplerian disks is much larger than amplitudes of the rest of two stable modes $A_0\sim\gamma\gg A_1\sim H_1 \sim \gamma ^{3/2}$.
 As a result, all terms in eqs. (\ref{amplitude})-(\ref{firstorder})  are of the same order
 in $\gamma$
 as
distinct from classical dispersive plasma media
 with $A_0\sim  A_1\sim H_1 \sim \gamma$,
  where the
 self-excitation cubic term  in eq. (\ref{amplitude}) is neglected as compared with the resonant quadratic term.

Equations (\ref{AC})-(\ref{firstorder}) are  inserted now  into the MHD equations, and solved order by order in $\gamma$. As expected, the lowest order reproduces the linear results
 for the eigenfunctions $\Psi_0(\eta)$, $\Psi_1(\eta)$ and $\Phi_1(\eta)$ in  relations (\ref{zeroharmonic})-(\ref{firstharmonicm}).
 The next order yields non homogeneous linear ordinary differential equations in $\eta$ that are subject  to zero boundary conditions at the disk edges for the following four  nonlinear  coupling functions: (i) zeroth MRI harmonics $\psi_{0,0}(\eta)$  and $\psi_{-1,1}(\eta)$; (ii) first AC harmonic   $\psi_{0,1}(\eta)$,  and (iii) first MS harmonic  $\phi_{0,1}(\eta)$.
 Since the  associate homogeneous   problems  for  $\psi_{0,0}(\eta)$,  $\psi_{-1,1}(\eta)$,    $\psi_{0,1}(\eta)$  and  $\phi_{0,1}(\eta)$  coincide with the eigenvalue problems for    $\Psi_{0}(\eta)$, $\Psi_{1}(\eta)$  and $\Phi_{1}(\eta)$, respectively, solutions of the corresponding non  homogeneous   problems have singularity. To avoid that singularities the four coupling coefficients ($\Gamma _{0,0}, \Gamma _{-1,1},  \Gamma _{0,1}, \Gamma _{1,0}$)  are introduced into eqs. (\ref{amplitude})-(\ref{firstorder})and are determined from the solvability conditions for the  non homogeneous problems for $\psi_{0,0}(\eta), \psi_{-1,1}(\eta), \psi_{0,1}(\eta), \phi_{0,1}(\eta)$.
The resulting values of ($\Gamma _{0,0}, \Gamma _{-1,1},  \Gamma _{0,1}, \Gamma _{1,0}$) are  exactly the same as for the strict resonance system
  \cite{Shtemler:2014uq}. The difference between the two systems is only in the exponential factors $e^{\pm i\delta\omega\, {\tilde \tau}}$ in eqs.  (\ref{amplitude})-(\ref{firstorder}) that describe the
    effect of
 small mismatch
  on the evolution of  resonant  triads.

Finally, by appropriately rescaling the amplitudes eqs. (\ref{amplitude})-(\ref{firstorder}) may be recast in the following form:
\begin{eqnarray}
\frac{d^2 a_0}{d{\tilde \tau}^2}=a_0-a_0^3+\sigma _0 \alpha^2[a_1^*h_1e^{-i\delta\omega\, {\tilde \tau}}+c.c]\label{rescaleda},\\
\frac{da_1}{d{\tilde \tau}}=i\alpha a_0h_1e^{-i\delta\omega\, {\tilde \tau}}, \; \frac{dh_1}{d{\tilde \tau}}=i\sigma _1 \alpha a_0a_1e^{i\delta\omega\, {\tilde \tau}},
\label{rescaledah}
\end{eqnarray}
where $a_0\sim a_1\sim h_1\sim\gamma^0$ are the rescaled amplitudes
$A_0$, $A_1$, $H_1$,
 respectively;  $\alpha = \sqrt{|\Gamma _{0,1}\Gamma _{1,0}/\Gamma _{0,0}|}$, $\sigma _0 =sign(\Gamma _{0,1}\Gamma _{-1,1})$, and $\sigma _1 =sign(\Gamma _{0,1}\Gamma _{1,0})$.
Equations (\ref{rescaleda})-(\ref{rescaledah}) constitute the dynamic  system that is investigated in the next section. Each interacting triad is thus characterized by a set of three coefficients $\alpha, \sigma _0$ and $\sigma _1$ which determines the nature of its dynamic  evolution.
The  MRI amplitude in thin disks
is governed by a Duffing equation of   second order
in contrast to the first order  equation for
 classical dispersive plasma media and fluids.

\subsection{\label{sec:level2C}C. The generalized Manley-Rowe relations}
A convenient representation of the amplitudes of the AC and MS modes is given by
$a_1({\tilde \tau})=\rho _a({\tilde \tau})e^{-i\varphi _a({\tilde \tau})}, h_1({\tilde \tau})=\rho _h({\tilde \tau})e^{-i\varphi _h({\tilde \tau})}$
(recall that $a_0$ is real-valued). In terms of these variables
equations (\ref{rescaleda})-(\ref{rescaledah}) give rise to the following two constants of motion:
\begin{eqnarray}
E_0&=&\frac{1}{2}\Bigl (\frac{da_0}{d{\tilde \tau}}\Bigr )^2-\frac{1}{2}\Bigl (a_0^2-\frac{1}{2}a_0^4\Bigr )+{\cal E} \label{manley1},\\
E_1&=&\frac{1}{2}\rho _h^2 +\sigma _1\frac{1}{2}\rho _a^2.
\label{manley2}
\end{eqnarray}
Here ${\cal E}=\frac{1}{2}\sigma _0\alpha \delta\omega\,(\sigma _1\rho _h^2-\rho _a^2)-2\sigma _0\alpha^2a_0\rho _a\rho _h cos\varphi$,
$\varphi =\varphi _a-\varphi _h -\delta\omega\, {\tilde \tau}$, and $E_0$ and $E_1$ are constants.
Equations
(\ref{manley1}) and (\ref{manley2})
 are
 the  Manley-Rowe type relations that are generalized
 for the amplitude of the MRI mode governed by  the  Duffing equation of the second order.

\section{\label{sec:level4}IV. Solutions for triad amplitudes}
The solutions of eqs.
(\ref{rescaleda})-(\ref{rescaledah}) exhibit much more complex behavior than their strict-resonance counterparts. As is demonstrated in \cite{Shtemler:2013qv}, \cite{Shtemler:2014uq} resonant triads with $\delta\omega=0$ are divided into stable ($\sigma _1=+1$) and unstable ($\sigma _1=-1$) triads. In the unstable case the MRI mode saturates by nonlinearly coupling to AC and MS modes that grow exponentially,
  while all the three amplitudes within stable triads are bounded and exhibit oscillatory behavior.
 As observed in Table 1 triads are stable ($S_{k,l}$) if the role of the stable AC  wave is played by a fast AC mode, and unstable  ($U_{k,l}$)   for slow  AC   waves.
This classification is inherited from the strict resonance case,
 however,  the exponential growth of  amplitudes  is  changed by explosive
  one
   for detuned resonance triads.
Table 1 presents the coefficients of systems
 (\ref{amplitude})-(\ref{firstorder})  and (\ref{rescaleda})-(\ref{rescaledah})
 for seven (out of infinitely many)  triads composed of MS, fast or slow AC modes, and  a background MRI mode ($k=1,l=-1$) at its threshold beta $\beta_1=2/3$.
  The triad stability criteria that emerge from Table 1 are that $\sigma_{1}=1>0$ is necessary and sufficient while $\sigma_{1}=-1<0$ is necessary for
  EMRI
 (also  for unstable triads  $\sigma_{0}=-1$).
    This will be demonstrated below explicitly  and numerically.
 \begin{table*}
 \centering
 \begin{minipage}{165mm}
\caption{
Eigen-frequencies $\omega_{k,l}(\beta_1)$ of the  AC modes and nonlinear  coupling coefficients of
systems (\ref{amplitude})-(\ref{firstorder})  and (\ref{rescaleda})-(\ref{rescaledah})
for resonant triads composed of the stable MS and fast ($l=1$) or slow $l=1$) AC modes ($k=1,2,3,\dots$), and  the background MRI mode ($k=1,l=-1$) at its threshold beta $\beta_1=2/3$;  $k $ is the axial wavenumber;  $l$ is the indicator of fast or slow  AC mode.   $S_{k,l}$ and $U_{k,l}$ are stable and unstable  triads  for fast and slow  AC modes.
 The stabilizing  non-resonant coupling coefficient  $\Gamma_{0,0} =-27/35$ \cite{Liverts:2012fr}.
     }
       \begin{tabular}{@{}cccccccc@{}}
         Triad
       & Stable & Unstable   & Stable & Unstable
            & Stable & Unstable  & Stable \\
  $\{S_{k,l}$,  $U_{k,l}\}$    & $S_{1,1}$ & $U_{2,-1}$    & $S_{2,1}$  & $U_{3,-1}$
            & $S_{3,1}$  & $U_{4,-1}$   & $S_{4,1}$
 \\
        \hline
   $k $          & $1$ &   $2$ & $2$ & $3$ & $3$ & $4$ & $4$
 \\
        \hline
$l$    & $1$ &   $-1$ & $1$ & $-1$ & $1$ & $-1$ & $1$
    \\
        \hline
     $\beta_{k}$
     & $2/3$ & $2$   &   $2$     & $4$
       & $4$    & $15$    & $15$
                   \\
        \hline
      $\omega_{k,l}(\beta_1)$
       & $\sqrt{7}$ & $\approx \sqrt{7/2}$   &   $\approx \sqrt{31/2}$     & $\sqrt{10}$
           & $\sqrt{27}$    & $\approx \sqrt{39/2}$    & $\approx \sqrt{83/2}$
                                          \\
           \hline
        $\Gamma_{1,0}$
       & $-0.68$ & $-3.40$   &$-5.06$    & $2.01$
          & $1.81$   & $16.46$   & $13.81$
                                        \\
                       \hline
        $\Gamma_{0,1}$
       & $-0.20$ & $94.97$    & $-0.03$    & $-4.40$    & $0.11$    & $-0.44$    & $0.011$
       \\
         \hline
      $\Gamma_{-1,1}$
       & $-0.14$ & $-2.16$   & $0.026$     & $3.78$
            & $0.66$     & $1.66$     & $-0.01$
          \\
                    \hline
      $\alpha=\sqrt{|\frac{\Gamma_{0,1}\Gamma_{1,0}}{\Gamma_{0,0}}|}$
       & $0.42$ & $20.5$   & $0.44$    & $3.4$
          & $0.51$   & $3.07$   & $0.44$
        \\
                    \hline
           $\sigma_{0}=sign(\Gamma_{0,1}\Gamma_{-1,1})$
      &$+1$ & $-1$  & $-1$    & $-1$
          & $+1$   & $-1$  & $-1$
             \\
                   \hline
          $\sigma_{1}=sign(\Gamma_{1,0}\Gamma_{0,1})$
       & $+1$ & $-1$   & $+1$    & $-1$
          & $+1$   & $-1$   & $+1$
                  \end{tabular}
\end{minipage}
\end{table*}

\subsection{\label{sec:level2A} A. Unstable triads ($\sigma_1=-1$)}
Solving eqs. (\ref{rescaleda})-(\ref{rescaledah}) numerically indicate that unlike the strict-resonance
case,
 all three modes may grow explosively in time, namely,  the solutions tend to infinity within a finite time, which is termed the explosion time  and denoted by ${\tilde \tau}_e$. To simplify the calculations, the particular case $\rho ({\tilde \tau})\equiv\rho _a({\tilde \tau})=\rho _h({\tilde \tau})$ is considered:
\begin{eqnarray}
\frac{d^2 a_0}{d{\tilde \tau}^2}=a_0-a_0^3+2\sigma_0 \alpha^2\rho^2\cos\phi,\,\,\,\,\,\,\,\,\,\,\,\,\,\,\,\,\,\,\,\,\,\,\,\,\,\,\nonumber\\
 \frac{d\rho}{d{\tilde \tau}}=-\alpha a_0\rho\sin\phi, \;
\frac{d\phi}{d{\tilde \tau}}=-2\alpha a_0\cos\phi- \delta\omega.
\label{simplified}
\end{eqnarray}
As follows from the forced Duffing  equation (\ref{simplified}) for the MRI amplitude
 the force term  exhibits temporal oscillations  instead of being a  constant in the strict resonance case.
  Numerical solutions demonstrate that for such triads, as
${\tilde \tau} \rightarrow {\tilde \tau}_e$ ,  $a_0$ and
$ \rho \rightarrow \infty$, while $\varphi \rightarrow (2M+1)\pi/2+\varphi _e$ where $M$ is an integer and $\varphi _e \rightarrow 0$,  such that $a_0\varphi _e$ remains finite. These observations lead to the ansatz $a_0({\tilde \tau})\varphi _e({\tilde \tau})=Const$ that results in the following asymptotic  expressions
 close to ${\tilde \tau}_e$:
\begin{equation}
{\hat a}_0\simeq \frac{K_0}{1-{\hat \tau}},\;\;\;\; {\hat \rho}\simeq \frac{K_1}{(1-{\hat \tau})^2}, \;\;\;\; {\hat \varphi} \simeq K_2(1-{\hat \tau}),
\label{explosive}
\end{equation}
where
$K_0=2  (-1)^{M+1}$, $K_1^2= sign(\delta\omega) (6+24/\alpha^2)$, $K_2=-1/3$; ${\hat \tau}\equiv {\tilde \tau}/{\tilde \tau}_e$;  ${\hat a}_0$, ${\hat \rho}$ and ${\hat \varphi}$ are the  normalized amplitudes and phase, respectively.
 It is evident that under the ansatz
  explosive solutions exist only for $\delta\omega>0$.
 With that restriction
 asymptotic  expressions (\ref{explosive})  in the  blow up vicinity  are \textit{universal} in the sense that they are independent of both $\delta \omega$ and $\tilde \tau_e$.
Thus ${\tilde \tau}_e$ cannot be calculated by the analysis that leads to eqs. (\ref{explosive}) and is
determined by the initial conditions.
Although  $\delta \omega>0$  in eqs. (\ref{explosive}),
 numerical solutions demonstrate  that explosive solutions exist also for $\delta\omega<0$, for which $a_0({\tilde \tau})\varphi _e({\tilde \tau})$ is not a constant near ${\tilde \tau}_e$ as it required in eqs. (\ref{explosive}).

Asymptotic expansions of eqs. (\ref{rescaleda})-(\ref{rescaledah}) or   (\ref{simplified}) in the artificial large parameter $\alpha$ yield that   $\alpha\tilde{\tau}$ and $\delta\omega/\alpha$ are the natural scales for slow time and frequency mismatch.
Thus neglecting the first two terms in the righthand side of the first of  eqs. (\ref{simplified}) yields   $a_0$, $\rho$ and $\phi$ for unstable triads that are uniformly valid up to $\alpha^{-2}$  in the whole range  of  $\alpha$ in Table 1.
In particular this implies that the principal unstable triad, $U_{2,-1}$,
 has  the biggest among other unstable triads   value of $\alpha=20.5$, and hence
 the shortest ${\tilde \tau}_e$.
    In this sense  that  triad is deemed the most unstable,
 and hence is the mostly used for the numerical demonstration in the following.

Figure \ref{Fig1} presents a comparison between the numerical solution of eqs. (\ref{simplified})  and the universal explosive solutions (\ref{explosive}).
It is seen that after an initial transient period, as time approaches the explosive time, the two dashed lines, which represent the numerical solutions of eqs. (\ref{simplified}) for two different
 values of $\delta\omega /\alpha$,
 converge to the  full line, which represents solution of eqs. (\ref{explosive}).

\begin{figure}[h]
\centering
\includegraphics*[width=85mm,height=35mm]{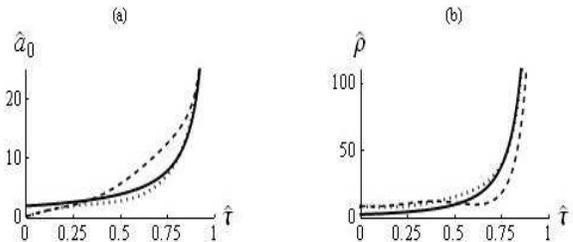}
\caption{
Comparison of the numerical solution of eqs. (\ref{simplified})  with universal explosive solutions (\ref{explosive}) for the principal unstable triad $U_{2,-1}$
($\alpha =20.5$).
Full line: solution of eqs. (\ref{explosive})  with  $\delta\omega /\alpha$=1. Lower and upper dashed lines: solution of eqs. (\ref{simplified}) with  $\delta\omega /\alpha$=1 and 2.5, respectively,  in the partial initial data ($\tilde\tau=0$: $a_0=0.05$, $da_0/d\tilde\tau=0.5\alpha$, $\rho=0.4$, $\phi=\pi/4$).
}
\label{Fig1}
\end{figure}

The   non-normalized amplitudes of the
 unstable triad, as obtained from the numerical solution
eqs. (\ref{simplified}), are depicted in Fig. \ref{Fig2}
 for typical value of the frequency mismatch $\delta\omega /\alpha$= 0.5.
 Note that the negative sign of $\sigma _1$
  enables the unbounded
growth of the amplitudes of the participating modes while keeping $E_1$ constant according to the Manley-Rowe relation (\ref{manley2}).
 It is interesting to note that while
$A_0\sim\gamma/(\tilde \tau-\tilde \tau_e)$ and
$A_1\sim H_1\sim\gamma^{3/2}/(\tilde \tau-\tilde \tau_e)^{2}$ the amplitude of MRI, initially  much larger,  grows much slower in the blow up vicinity  than the other of two modes,
 the amplitudes become comparable
 when $(\tilde \tau-\tilde \tau_e)\sim\gamma^{1/2}$.

\begin{figure}[!]
\centering
\includegraphics*[width=85mm,height=35mm]{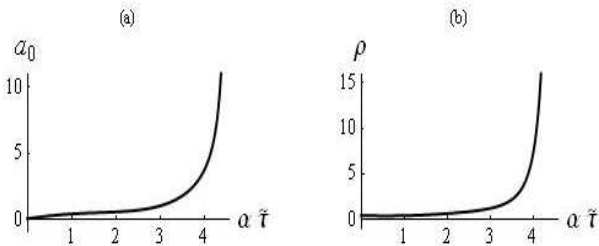}
\caption{
Solution of eqs. (\ref{simplified})
 in the partial initial data ($\tilde\tau=0$: $a_0=0.05$, $da_0/d\tilde\tau=0.5\alpha$, $\rho=0.4$, $\phi=\pi/4$)
 for   the principal unstable triad $U_{2,-1}$
($\alpha =20.5$, $\delta\omega /\alpha=0.5$, $\alpha \tilde{\tau}_e \approx 4.5$).
}
\label{Fig2}
\end{figure}

   Since the blow-up will occur  earlier for the shortest ${\tilde \tau}_e$
its  value    is an effective  measure of the nonlinear instability.
 Since  strict resonance unstable triads ($\delta\omega\equiv0$) grow exponentially they have ${\tilde \tau}_e=\infty$, while  detuned resonance triads have rather bounded values ${\tilde \tau}_e$, when $\delta\omega$ approaches $\pm0$
 (Fig. \ref{Fig3}), except of the vicinity of the mismatch threshold value.
 Indeed,
 Fig. \ref{Fig3} for two different sets
of the initial data demonstrates that
 dependence of ${\tilde \tau}_e$ vs.  $\delta\omega$ is strongly influenced by the initial data.
 Although
   EMRI
  occurs for any  $\delta\omega>0$ with  the shortest value
${\tilde \tau}_e$ of about $0.2$
 for both sets of the initial data,
 for $\delta\omega<0$
  EMRI
  occurs only when  $\delta\omega$ does not exceed the threshold given by
   $|\delta\omega_{th}|/\alpha$ (that value significantly depends on initial data),
  otherwise the amplitudes of the triad  oscillate. This also implies that in the limit of vanishing frequency-mismatch the triad's amplitudes do not convergence to those in the strict resonance case.

\begin{figure}[!]
\centering
\includegraphics*[width=75mm,height=45mm]{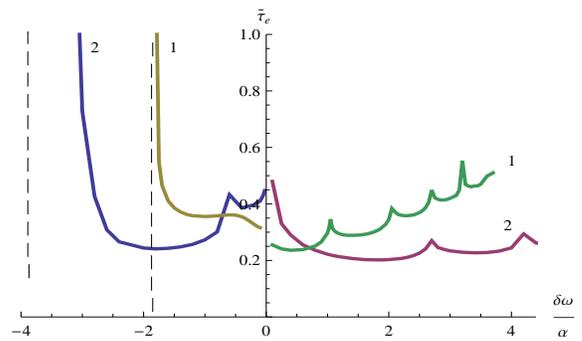}
\caption{
Dependence of ${\tilde \tau}_e$ vs.  $\delta\omega/\alpha$.
Solution of eqs. (\ref{simplified})
 for   the principal unstable triad $U_{2,-1}$
($\alpha=20.5$), in two different sets of the  initial data, 1 and 2:
  1.-$\tilde\tau=0$: $a_0=0.05$, $da_0/d\tilde\tau=0.5\alpha$, $\rho=0.4$, $\phi=\pi/4$;
  2.-$\tilde\tau=0$: $a_0=0.05$, $da_0/d\tilde\tau=0$, $\rho=0.4$, $\phi=\pi/4$.
The vertical dashed straightlines depict the  mismatch thresholds for initial data 1. and 2.:  $\delta\omega_{th}/\alpha\approx -1.85$ and  $\delta\omega_{th}/\alpha\approx -3.9$, respectively.
}
\label{Fig3}
\end{figure}

\subsection{\label{sec:level2B} B. Stable triads ($\sigma_1=+1$)}
The amplitudes of the  stable triad
 $S_{1,1}$
 remain bounded at all times and exhibit irregular periodic oscillations.
 This may be seen in Fig. \ref{Fig4} where a saturation of the MRI is demonstrated
for  numerical solutions of eqs. (\ref{rescaleda})-(\ref{rescaledah}).
\begin{figure}[!]
\centering
\includegraphics*[scale=1.25]{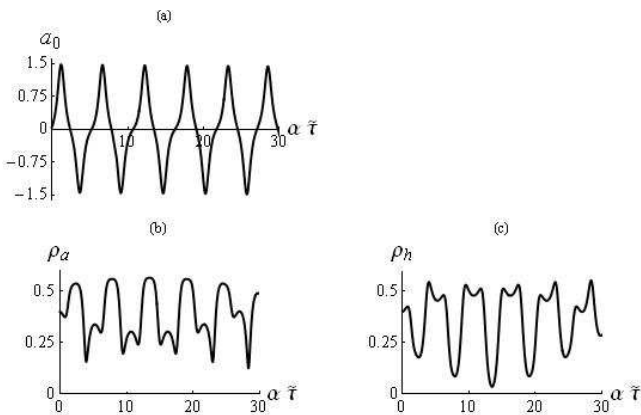}
\caption{Solution of eqs. (\ref{rescaleda})-(\ref{rescaledah})
in the partial initial data  ($\tilde\tau=0$: $a_0=0.05$, $da_0/d\tilde\tau=0.5\alpha$, $a_1=h_1=0.4$) for the stable triad $S_{1,1}$
($\alpha =0.42$, $\delta\omega /\alpha=1$).
   }
\label{Fig4}
\end{figure}

\section{\label{sec:level5}V. Discussion and Summary}
Is is shown  that at resonance conditions the weakly-nonlinear evolution  of  unstable triads formed by the background MRI mode (first unstable slow AC mode) together  with the associated  two linearly stable MS and another slow AC modes can give rise to EMRI  in the vicinity of its linear stability threshold.

The triad amplitude relations for thin Keplerian disks   are similar to those for other dispersive plasma media  \cite{Fukai:1969} \nocite{Dum and Ott:1971} \nocite{Davydova et al:1978} --\cite{Pavlenko:1978}, but exhibit some specific features.
 As was first noted in \cite{Dum and Ott:1971}  for beam plasmas (see also detailed studies in \cite{Davydova et al:1978} and \cite{Pavlenko:1978})
 zero-energy mode can excite EI  in the vicinity of the linear stability threshold.
 Indeed, energy of the linear MRI is zero as was shown in  \cite{Khalzov et al:2008}.
 As indicated in \cite{Dum and Ott:1971} \nocite{Davydova et al:1978} --\cite{Pavlenko:1978}, this results in the appearance of the second time derivative in the  amplitude equation for zero-energy mode (see also \cite{Liverts:2012fr}).
 Along with this there are some important distinctions  between the two systems, among which the non-zero frequency mismatch  is obligatory for EI in thin-disk plasmas, but it is not necessary for beam plasmas.
In addition,  the  self-similarity laws for amplitudes of the triad in time in the blow-up point for beam plasmas \cite{Davydova et al:1978}  differ from those for thin-disk plasmas.
 In-spite of those differences  the qualitative agreement between both systems is obvious.
 Also  EMRI in thin disks either becomes less degenerous due to growth of the explosive time with increasing   $|\Delta\omega|$, or, as in \cite{Fukai:1969},   saturated by turning into bounded oscillations  for negative $\Delta\omega$ less than a threshold value (Fig. \ref{Fig3}).
 Besides, as opposite to
    beam-plasma systems in \cite{Fukai:1969} \nocite{Dum and Ott:1971} \nocite{Davydova et al:1978} --\cite{Pavlenko:1978} (for which $\Delta\omega$   assumed to be much smaller than the non-zero  eigenfrequencies of all three interacting modes),
    for thin-disk plasmas the MRI unstable eigenmode  has the zero-real frequency  near the linear stability threshold, while $\Delta\omega$  should be  of the order of small growth rate $\gamma$
of the MRI and much smaller than real eigenfrequencies of the rest two   waves.
  Finally note that EMRI in thin disks
   is driven by the positive feedback between the MRI mode and the rest two  waves.
   Under resonance conditions the positive feedback is provided by the force term in the Duffing equation for the MRI amplitude, where the force term
 grows  and the MRI amplitude with it  when amplitudes of the associated waves rise.
     In turn the amplitudes of the associated waves rise with  amplitude of the MRI mode. However for strict resonance conditions
   the positive feedback is degenerated. Indeed, amplitudes of the associated
  waves grow exponentially,  but the feedback leaves  the amplitude of the axisymmetric MRI mode bounded    because of the force term  constancy in time \cite{Shtemler:2014uq}.
   This degeneration of the positive feedback at $\delta \omega=0$ means that the presence of a small non-zero mismatch does not merely lead to small
 deviations from the strict-resonance case, but may change the behavior of the system in a fundamental way.

 As physically evident  EMRI should be saturated with time.
 Nonlinear terms with  higher order harmonics in time neglected in the current  approximation have been taken into account in a number of earlier studies in order to saturate EI (\cite {Weiland} and references therein). However, this has been found ineffective by Mahoney \cite{Mahoney:1995} which shows that all higher order harmonics are of the same order and  should be involved simultaneously   in the blow-up   vicinity that rather requires solution of the compete nonlinear problem.
       It is suggested that the analytical results obtained in the present study will be useful for elucidating physical irregularities of  MRI.
    It is conjectured that
the  EMRI    describes  an
 intermediate stage of the MRI nonlinear evolution  during which a significant energy transfer   takes place.
 That stage of the nonlinear interaction between
  the three resonant
   modes occurs
    before the blow-up and leads to the break-down of the present model.  Direct numerical simulations
     can resolve the evolution beyond blow-up.

\section*{Acknowledgments}
\acknowledgments
 This work was supported by grant number 366/15 of the Israel Science Foundation.

\end{document}